\documentclass{article}

\usepackage[final]{Arxiv_v1}

\usepackage[utf8]{inputenc} 
\usepackage[T1]{fontenc}    
\usepackage{hyperref}       
\usepackage{url}            
\usepackage{booktabs}       
\usepackage{amsfonts}       
\usepackage{nicefrac}       
\usepackage{microtype}      

\usepackage[pdftex]{graphicx}
\usepackage{amssymb,amsfonts,amsmath}
\usepackage{color}

\definecolor{white}{rgb}{1,1,1}
\definecolor{blue}{rgb}{0.1,0.1,1}
\definecolor{cyan}{rgb}{0.55,0.6,0.8}  
\definecolor{red}{rgb}{0.8,0.2,0.2}  
\definecolor{green}{rgb}{0,0.8,0} 
\definecolor{purple}{rgb}{0.6,0,1}
\definecolor{orange}{rgb}{1,0.6,0.3}

\newcommand{\blue}{\color{black}}
\newcommand{\cyan}{\color{black}}
\newcommand{\red}{\color{black}} 
\newcommand{\green}{\color{black}}
\newcommand{\purple}{\color{black}}

\title{Gradient Descent  for Spiking Neural Networks}

\author{
Dongsung~Huh
  \\
  Salk Institute\\
  La Jolla, CA 92037 \\
  \texttt{huh@snl.salk.edu} \\
   \And
  Terrence J. Sejnowski \\
  Salk Institute\\
  La Jolla, CA 92037 \\
  \texttt{terry@salk.edu}\\}

\begin{document}

\maketitle

\begin{abstract}

Much of studies 
on neural computation are based on network models of static neurons that produce analog output, 
despite the fact that 
information processing in the brain is predominantly carried out by 
dynamic neurons that produce discrete pulses called  
spikes. 
{\blue Research in spike-based computation has been impeded by} the lack of efficient supervised learning algorithm for spiking networks. 
{\blue Here, we present a gradient descent method for optimizing spiking network models by introducing a differentiable formulation of spiking networks and deriving the exact gradient calculation.}
For demonstration, we trained recurrent spiking networks on two dynamic tasks:
one that requires optimizing fast ($\approx$ millisecond) 
spike-based interactions for efficient encoding of information, 
and a delayed-memory XOR task 
over extended duration 
($\approx$ second).
The results show that our method indeed optimizes the spiking network dynamics on the time scale of individual spikes as well as the behavioral time scales. 
%
%
{\blue In conclusion, our {\green result} offers a general purpose supervised learning algorithm for spiking neural networks, 
{\red thus advancing further investigations} on spike-based computation. 
}



\end{abstract}

\section{Introduction}

The brain executes highly dynamic computation over multiple time-scales:
Individual neurons exhibit spiking dynamics of millisecond resolution and 
the 
recurrent network connections 
produce slower dynamics on the order of seconds to minutes.
How the brain organizes  the spiking neuron dynamics to form the basis for computation is a central problem in neuroscience.  
Nonetheless, 
%
most analysis and modeling of neural computation 
{\green assume} 
static neuron models that produce analog output,
{\blue called rate-based neurons}
{\purple \cite{mante2013context,yamins2014performance}}.
These simplified models are compatible with the advanced {\green tools} from the deep learning field
that can efficiently optimize large scale network models  to perform complex computational tasks 
{\purple \cite{lecun2015deep}}.
However, such {\blue rate-based neural network} models fail to describe 
the fast dynamics of spike-based computation in the brain.


The main difficulty in optimizing spiking neural networks 
stems from the discrete, all-or-none nature of spikes: 
A spiking neuron generates a brief spike output when its membrane voltage crosses the threshold, 
and silent at other times. 
This non-differentiable behavior is incompatible with the standard, gradient-based supervised learning methods.
Consequently, previous learning methods 
for spiking neural networks {\blue explored various ways} to circumvent the non-differentiability problem.

SpikeProp {\purple \cite{bohte2002error}}
considered the spike times of neurons as state variables and used the differentiable relationship between the input and the output spike times to minimize the difference between the actual and the desired output spike times. 
However, the creation and deletion of spikes are non-differentiable, so the number of output spikes must be pre-specified. 
Memmesheimer et al {\purple \cite{memmesheimer2014learning}} considered the problem of generating spikes at desired times and remaining silent at other times as a binary classification problem and applied the perceptron learning rule {\red (See also {\purple\cite{gutig2006tempotron}})}.
Pfister et al {\purple \cite{pfister2006optimal}} considered stochastic spiking neurons  
and maximized the smooth likelihood function of the neurons spiking at desired times. 
More biologically inspired methods based on spike-time-dependent-plasticity (STDP) have also been proposed
{\purple \cite{florian2007reinforcement,izhikevich2007solving,legenstein2008learning,ponulak2010supervised}}.
%
All of these methods, however, require target spiking activity of individual neurons at desired times, 
which significantly limit their range of applicability.

Alternative methods have also been proposed:
Instead of directly optimizing a spiking network,
these methods optimize a network of static analog neurons 
and replicate the optimized solution 
with a spiking network. 
Hunsberger and Eliasmith {\purple \cite{hunsberger2015spiking}} used analog units 
that closely approximated the firing rate of individual spiking neurons.
Instead of replicating individual neuron's dynamics,
Abbott et al  {\purple \cite{abbott2016building}} proposed
replicating the entire network dynamics 
using recently developed methods from {\it predictive coding}  {\purple \cite{deneve2016efficient}}.
Although these approaches are applicable 
to a wider range of problems,
the replicated spiking networks can only mimic the solutions of the 
{\blue rate-based} networks, 
rather than exploring the larger space of spike-time based solutions.


In this paper, 
we introduce a novel, differentiable formulation of spiking neural networks and derive the exact gradient calculation for gradient based optimization. 
This method optimizes  the recurrent network dynamics on the time scale of individual spikes  
for general supervised learning problems.

\section{Methods} 


\subsection{Differentiable synapse model}

In spiking networks,  transmission of neural activity is mediated by synaptic current.
Most models describe the synaptic current dynamics as 
a linear filter process which instantly activates 
when the presynaptic membrane voltage $v$ crosses a threshold:
{\it e.g.}, 
\begin{align}
	\tau \dot{s} &  = -  s  +  \sum_k  \delta(t-t_k).  
	\label{eq:syn_dyn_old}
\end{align}
where  $\delta(\cdot)$ is the Dirac-delta function, and $t_k$  denotes the time of threshold-crossing. 
Such threshold-triggered dynamics generates discrete, all-or-none responses of synaptic current, 
which is non-differentiable. 

Here, we replace the threshold with a gate function $g(v)$: 
a non-negative ($g \geq 0 $),  
unit integral  ($\int g ~ dv = 1$) function  with narrow support%
\footnote{Support of a function $g : X \to \mathbb{R}$ is the subset of the domain $X$ where $g(x)$ is non-zero.}, 
which we call the active zone.
This allows the synaptic current to be activated in a gradual manner throughout the active zone. 
The corresponding synaptic current dynamics is
\begin{align}
	\tau \dot{s} &  = - s  + g \dot{v} ,
	\label{eq:syn_dyn_new}	
\end{align}
where $\dot{v}$ is the time derivative of the pre-synaptic membrane voltage.
The $\dot{v}$ term is required for 
the dimensional consistency between eq~\eqref{eq:syn_dyn_old} and \eqref{eq:syn_dyn_new}: %
The $g\dot{v}$ term  
has the same $[\text{time}]^{-1}$ dimension as the Dirac-delta impulses of eq~\eqref{eq:syn_dyn_old}, 
since the gate function has the dimension $[\text{voltage}]^{-1}$ and $\dot{v}$ has the dimension $[\text{voltage}][\text{time}]^{-1}$. 
Hence, the time integral of synaptic current, {\it i.e.} charge, is a dimensionless quantity. 
Consequently, 
a depolarization event beyond the active zone induces a constant amount of total charge
regardless of the time scale of depolarization,
since 
$$ \int s~dt =  \int   g \dot{v}  ~ dt =  \int  g ~ dv =  1 . $$

Therefore, eq~\eqref{eq:syn_dyn_new} 
generalizes the threshold-triggered synapse model  
while preserving the fundamental %
property of spiking neurons: %
{\it i.e.} all supra-threshold 
depolarizations induce the same amount of synaptic responses regardless of the depolarization rate
(Figure 1A,B). 
Depolarizations 
below the active zone induce no synaptic responses (Figure 1E),
and depolarizations 
within the active zone induce graded responses 
(Figure 1C,D).
This contrasts with the threshold-triggered synaptic dynamics, 
which causes 
abrupt, non-differentiable change of response 
at the threshold
(Figure 1, dashed lines). 

Note that the $ g \dot{v} $ term reduces to the Dirac-delta impulses in the zero-width limit of the active zone, 
which reduces eq~\eqref{eq:syn_dyn_new}  back to the threshold-triggered synapse model  eq~\eqref{eq:syn_dyn_old}.

The gate function, without the $\dot{v}$ term, was previous used as a differentiable model of synaptic connection {\purple \cite{lajoie2013chaos}}.
In such a model, however, 
a spike event delivers varying amount of charge depending on the depolarization rate: 
the slower the presynaptic depolarization, 
the greater the amount of charge delivered to the post-synaptic targets. 

\begin{figure} 
  \centering
   \includegraphics[width=0.99\textwidth]{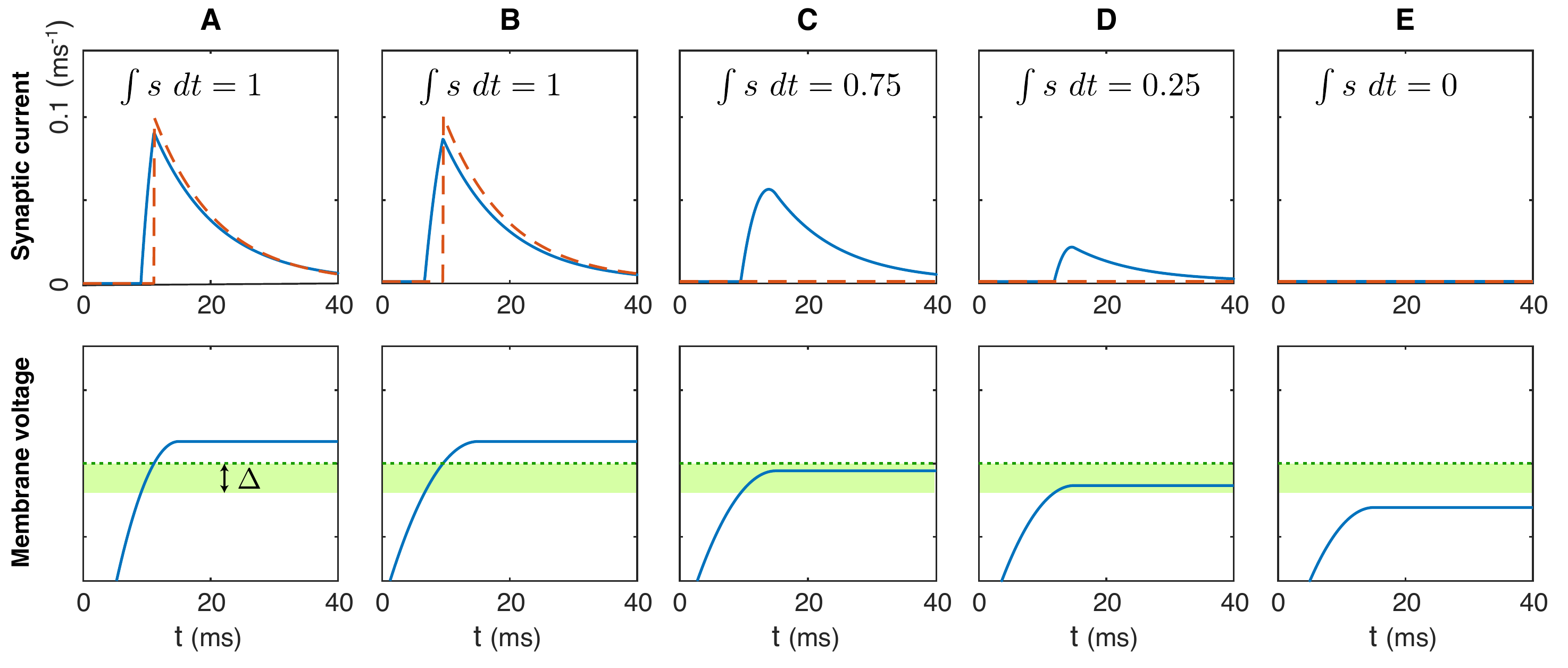}
  \caption{
Differentiability of synaptic current dynamics:
The synaptic current traces from eq~\eqref{eq:syn_dyn_new} (solid lines, upper panels)
are shown with the corresponding membrane voltage traces (lower panels).
Here, the gate function is $g = 1/\Delta $ within the active zone 
of width $\Delta$ (shaded area, lower panels){\red ;} $g = 0$ otherwise. 
(A,B) The pre-synaptic membrane voltage depolarizes beyond the active zone. 
Despite the different rates of depolarization, 
both events incur the same amount of charge in the synaptic activity: $\int s ~ dt = 1$.
(C,D,E) Graded synaptic activity due to insufficient depolarization levels that do not exceed the active zone.
The threshold-triggered synaptic dynamics in eq~\eqref{eq:syn_dyn_old} is also shown for comparison
(red dashed lines, upper panels). 
The effect of voltage reset is ignored for the purpose of illustration. 
$\tau = 10$ ms. 
}
\end{figure}

\subsection{Network model}

To complete the input-output dynamics of a spiking neuron,
the synaptic current dynamics 
must be coupled with 
the presynaptic neuron's internal state dynamics.  
For simplicity, we consider differentiable neural dynamics that depend only on the the membrane voltage and 
the input current:  
\begin{align}
	\dot{v} & = f(v, I) .
	\label{eq:memV_dyn}
\end{align}

The dynamics of an interconnected network of neurons can then be constructed by linking 
the {\green dynamics of individual neurons and synapses} 
eq~(\ref{eq:syn_dyn_new},\ref{eq:memV_dyn}) %
through the input current vector:  
\begin{align}
	\vec{I} = W \vec{s} + U  \vec{i} + \vec{I}_o, 
	\label{eq:I}
\end{align}
where $W$ is the recurrent connectivity weight matrix, 
$U$ is the input weight matrix,  $\vec{i} (t)$ is the input signal for the network, 
and $\vec{I}_o$ is the tonic current. 
Note that 
this formulation describes general, fully connected networks;
specific network structures can be imposed by constraining the connectivity:
{\it e.g.} triangular matrix structure $W$ for  multi-layer feedforward networks.  

Lastly, we define the output of the network
as the linear readout of the synaptic current:
\begin{align*}
	\vec{o}(t) = O \vec{s}(t),
\end{align*}
where $O$ is the readout matrix.

The network parameters $W$, $U$, $O$, $\vec{I}_o$  {\blue will be} optimized to minimize the total cost,
$C \equiv \int l (t) ~ dt $,
where $l$ is the cost function that evaluates the performance of network output for given task.

\subsection{Gradient calculation} 

The above spiking neural network model can be optimized via gradient descent. %
In general, the exact gradient of a dynamical system can be calculated using 
either Pontryagin's minimum principle {\purple \cite{pontryagin1962mathematical}}, also known as backpropagation through time, 
or real-time recurrent learning, which yield identical results. 
We present the former approach here,
which scales better with network size, $\mathcal{O}(N^2)$ instead of $\mathcal{O}(N^3)$,
but the latter approach can also be straightforwardly implemented. 

Backpropagation through time for the spiking dynamics 
eq~(\ref{eq:syn_dyn_new},\ref{eq:memV_dyn}) {\green utilizes} 
the following backpropagating dynamics of adjoint state variables (See Supplementary Materials):
\begin{align}
	\label{eq:p_v}
	- \dot{p}_{v} 	& = f_v  {p}_{v}  - g   \dot{p}_{s} \\
	\label{eq:p_s}
        -  \tau   \dot{p}_{s} &  =  -  {p}_{s}  + \xi ,
\end{align}
where $f_v  \equiv \partial f/ \partial v$,
and $\xi$ is called the {\it error current}. 
For the recurrently connected network eq~\eqref{eq:I},  
the error current vector has the following expression
\begin{align}
	\vec{\xi} =  W^{\intercal} \vec{({f}_I  {p}_{v})}   +  \vec \partial_{s} l , 
	\label{eq:xi}
\end{align}
which links 
the backpropagating dynamics eq~(\ref{eq:p_v},\ref{eq:p_s}) of individual neurons. 
Here, 
$f_I  \equiv \partial f/ \partial I$, 
$({f}_I  {p}_{v})_j \equiv f_{I_j}  {p}_{v_j}$, and 
$(\partial_{s} l)_j  \equiv \partial l/ \partial s_j$.

Interestingly, the coupling term of the backpropagating dynamics, $g \dot{p}_{s} $, has the same form as the coupling term $g \dot{v}$ of the forward-propagating dynamics. 
Thus, the same gating mechanism that mediates the spiked-based communication of signals
also controls the propagation of error in the same sparse, compressed manner.

Given the adjoint state vectors 
that satisfy eq~(\ref{eq:p_v},\ref{eq:p_s},\ref{eq:xi}),
the gradient of the total cost 
with respect to the network parameters 
can be calculated as
\begin{align*}
       \nabla_{W} C  &= \int \vec{({f}_I  {p}_{v})} ~ \vec{s}^{\intercal} ~ dt 
	\label{eq:gradient} \\  
       \nabla_{U} C  &= \int \vec{({f}_I  {p}_{v})} ~ \vec{i}^{\intercal} ~ dt \nonumber \\  
       \nabla_{I_o} C &=  \int \vec{({f}_I  {p}_{v})}  ~ dt \nonumber \\
       \nabla_{O} C   &  = \int \vec{\partial}_o l ~  \vec{s}^{\intercal}  ~ dt \nonumber 
\end{align*}
where 
$(\partial_o l)_j  \equiv \partial l/ \partial o_j$.
Note that the gradient calculation procedure 
involves multiplication between the presynaptic input source 
and the postsynaptic adjoint state $p_v$, 
which is driven by the $g \dot{p}_s$ term: {\it i.e.} the product of postsynaptic spike activity and temporal difference of error.
This {\green is analogous to}  
reward-modulated spike-time dependent plasticity (STDP) {\purple \cite{fremaux2015neuromodulated}}.

\section{Results} 

We demonstrate our method by training spiking networks on 
dynamic tasks that require information processing over time. 
Tasks are defined by the relationship between 
time-varying input-output signals, 
which are used as training examples.
We draw mini-batches of $\approx 50$ training examples from the signal distribution, 
calculate the gradient of the average total cost, 
and use stochastic gradient descent 
{\purple \cite{kingma2014adam}} for optimization. 

Here, we use a cost function $l$ that penalizes the readout error and the overall synaptic activity: 
\begin{align*}
l = \frac {\Vert   \vec{o} - \vec{o}_d  \Vert ^2 + \lambda  \Vert \vec{s} \Vert ^2}{2} , 
\end{align*}
where $\vec{o}_d (t)$ is the desired output, and $\lambda$ is a regularization parameter.

\begin{figure} 
  \centering
   \includegraphics[width=1\textwidth]{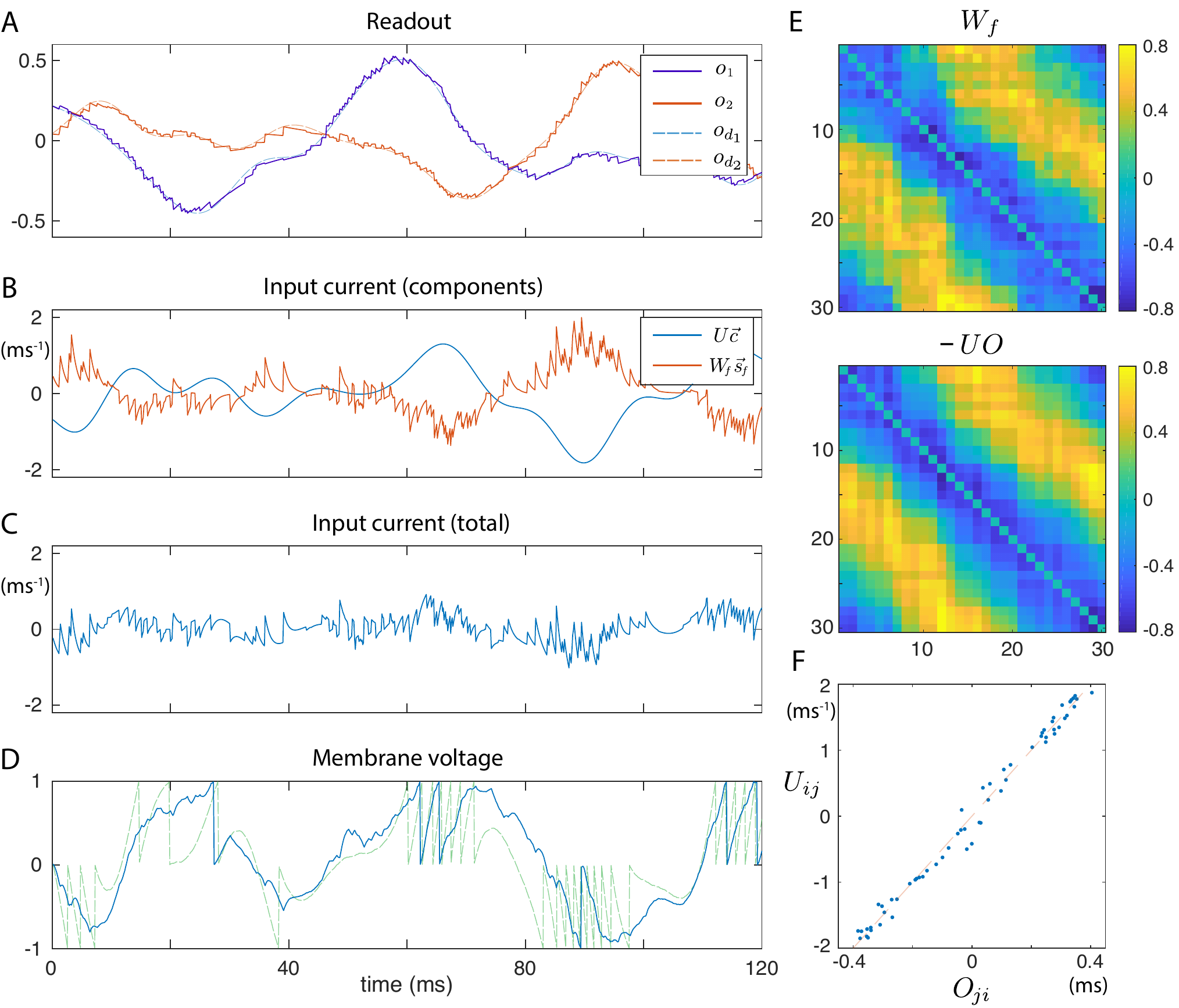}
  \caption{
Balanced dynamics of a spiking network trained for auto-encoding task.
(A) Readout signals: actual (solid), and desired (dashed). 
(B) Input current components into a single neuron: external input current ($U\vec{i}(t)$, blue),  and fast reccurent synaptic current ($W_f\vec{s}_f(t)$, red).
(C) Total input current into a single neuron ($U\vec{i}(t) + W_f\vec{s}_f(t)$).
(D) Single neuron membrane voltage traces: the actual voltage trace driven by both external input and fast reccurent synaptic current (solid, 6 spikes), and a virtual 
trace driven by external input only (dashed, 29 spikes).
(E) Fast recurrent weight: trained ($W_f$, up) and predicted ($- U O$, down).
Diagonal elements are set to zero to avoid self-excitation/inhibition.
(F) Readout weight $O$ vs input weight $U$.
}
\end{figure}

\subsection{Predictive Coding Task} 

We first consider {\it predictive coding} tasks 
{\purple \cite{deneve2016efficient,boerlin2013predictive}},
which optimize spike-based representations 
to accurately reproduce the input-ouput behavior of a linear dynamical system of 
full-rank input and output matrices.   
Analytical solutions for this class of problems can be obtained in the form of 
non-leaky integrate and fire (NIF) neural networks, 
although insignificant amount of leak current is often added.
The solutions also require the networks to be equipped with a set of instantaneously fast synapses, 
in addition to the {\it regular} synapses of finite time constant 
{\purple \cite{boerlin2013predictive}}.

The membrane voltage dynamics of a NIF neuron is given by 
$$f(v,I) = I . $$  
To ensure that the membrane voltage stays within a finite range, we impose two thresholds at 
 $v_{\theta+}=1$ and $v_{\theta-}=-1$, 
and the reset voltage at $v_\text{reset}=0$.%
\footnote{\blue The reset process, which may seem non-differentiable,  
does not influence the gradient calculation.}
For simplicity, 
we allow the $v_{\theta-}$ threshold to trigger negative synaptic responses, 
{\blue which can be turned off if desired.}  

We also introduce the additional fast synaptic current $\vec{s}_f $ 
proposed in {\purple \cite{deneve2016efficient,boerlin2013predictive}},  
which modifies the input current vector to be
$\vec{I} = W \vec{s} + W_f \vec{s}_f + U  \vec{i} + \vec{I}_o, $
where $W_f$ is the recurrent weight matrix associated with fast synapses.
However,
assigning zero time constant to the fast synapses  
often causes unstable dynamics, 
because it could lead to one spike immediately triggering more spikes in other neurons.
Here, we assign finite time constants for both types of synapses: 
$\tau_f = 1$ ms for fast synapses, and $\tau = 10$ ms for regular synapses.

Despite its simplicity, the predictive coding framework  reproduces important features of biological neural networks, 
such as the balance of excitatory and inhibitory inputs and efficient coding {\purple \cite{deneve2016efficient}}. 
Also, 
its analytical solutions provide a benchmark for  assessing results from optimization.

\paragraph{Auto-encoder task}

In the {\it auto-encoder} task, the desired output signal is a low-pass filtered version of the input signal: 
\begin{align*}
	\tau  \dot{\vec{o}}_d = -\vec{o}_d + \vec{i},      
\end{align*}
where $\tau$ is the synaptic time constant {\purple \cite{deneve2016efficient,boerlin2013predictive}}. 
The goal is to accurately represent the analog signals using least number of spikes.
We used a network of 30 NIF neurons, 
and $2$ input and output signals. 
Randomly generated sum-of-sinusoid signals with period $1200$ ms were used as the input. 
{\cyan $\lambda=0.1/N$ ms$^2$.}
{\cyan $\Delta=0.1$ was used for training, then set to zero for post-training simulations.}

The output of the trained network accurately tracks the desired output (Figure 2A).
Analysis of the simulation reveals that the network operates in a tightly balanced regime: 
The fast recurrent synaptic input, 
$W_f \vec{s}_f(t) $, provides {\green opposing} 
current that mostly cancels the input current from the external signal, $U \vec{i}(t) $,
such that the neuron generates a greatly reduced number of spike outputs (Figure 2B,C,D).
The network structure also shows close agreement to the prediction. 
The optimal input weight matrix is equal to the transpose of the readout matrix (up to a scale factor),
$U \propto O^\intercal$,
and the optimal fast recurrent weight is approximately the product of the input and readout weights,
$W_f \approx - U O$ , 
which are in close agreement with
{\purple \cite{deneve2016efficient,boerlin2013predictive,brendel2017learning}}. %
The {\it regular} recurrent connection is not needed for this task and hence $W$ was set to zero.  
Such network structures have been shown to maintain tight input balance and remove redundant spikes to encode the signals in most efficient manner: 
The representation error scales as $1/K$, where $K$ is the number of involved spikes,  
compared to the $1/\sqrt{K}$ error of encoding with independent Poisson spikes.

\begin{figure} 
  \centering
   \includegraphics[width=0.65\textwidth]{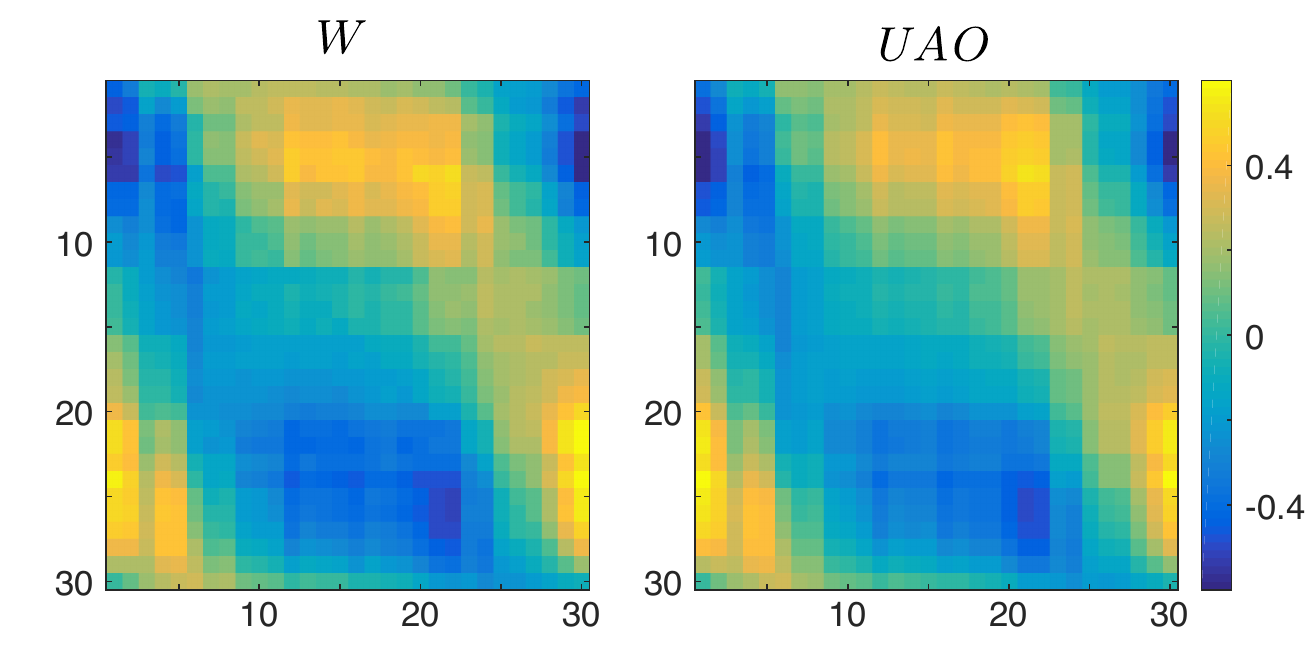}
  \caption{
Regular recurrent weight for the predictive coding task:  trained ($W$, left) and
 predicted ($U A O$, right).
}
\end{figure}

\paragraph{General predictive coding task} 

More generally, {\blue predictive coding tasks involve linear} dynamic relationships between the desired input-output signals of the following form: 
$$ \tau  \dot{\vec{o}}_d = -\vec{o}_d + A \vec{o}_d + \vec{i},$$ 
where $A$ is a constant matrix.
Here, we trained a spiking network of 30 NIF neurons
with $2$ input signals of sums-of-sinusoid and $A = [-0.7, 0.36; -2.3, -0.1]$, which strongly modulates the desired output signal dynamics.

Similar to the result shown in Figure 2, the trained network exhibits tightly balanced input current with the network output accurately tracking the desired output.
The optimal {\it regular} recurrent weight is approximately $W  \approx  U A O$  (Figure 3), which is also in close agreement with the prediction {\purple \cite{deneve2016efficient,boerlin2013predictive,brendel2017learning}}. 
The other network structures are similar to the case of auto-encoding task.

These results show that {\green our algorithm}  
accurately optimizes the millisecond time-scale interaction between neurons 
to find an efficient spike-time-based encoding scheme.
Moreover, it also shows that efficient coding can be robustly achieved without introducing instantaneously fast synapses, which were previously considered to be necessary. %

\subsection{Delayed-memory XOR task}

\begin{figure} 
  \centering
   \includegraphics[width=0.99\textwidth]{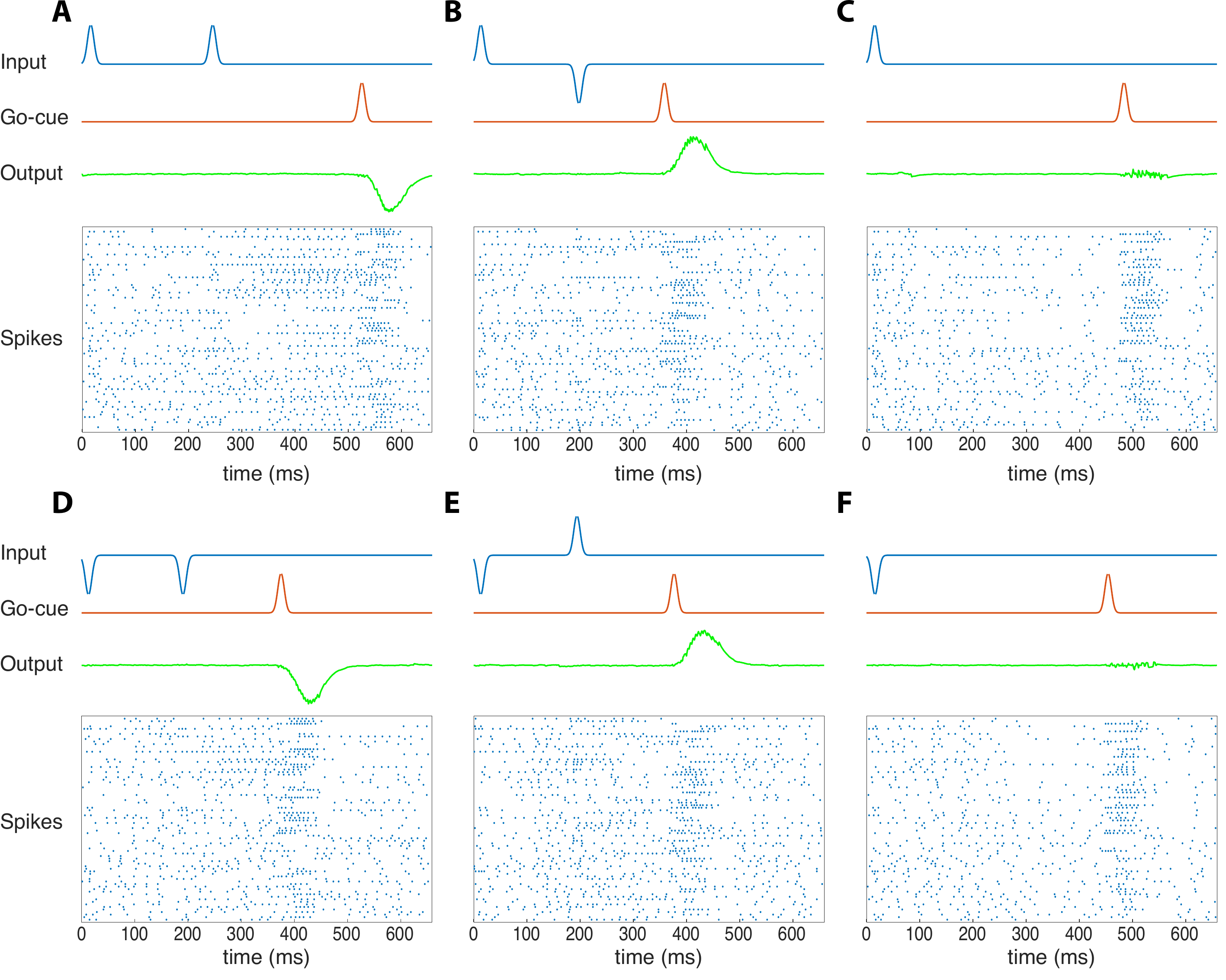}
  \caption{
Delayed-memory XOR task: Each panel shows  the single-trial input, go-cue, output traces, and spike raster of an optimized QIF neural network.  The y-axis of the raster plot is the neuron ID. 
{\red Note 
the similarity of the initial portion of spike patterns for trials of the same first input pulses 
(A,B,C   vs  D,E,F).
In contrast, the spike patterns after the go-cue signal are similar for trials of the same desired output pulses:
(A,D:  negative output), (B,E:  positive output), and (C,F:  null output).}
}

\end{figure}

A major challenge for spike-based computation is {\red in bridging} the wide divergence between the time-scales of behavior and spikes: How do {\red millisecond} spikes {\green perform  behaviorally relevant computations} on the order of seconds? 

Here, we consider a {\it delayed-memory XOR} task, 
which performs the {\it exclusive-or} (XOR) operation on the input history stored over extended duration. 
Specifically,  the network receives binary pulse signals, $+$ or $-$,  through an input channel and a go-cue through another channel. 
If the network receives two input pulses since the last go-cue signal, 
it should generate the XOR output pulse on the next go-cue: 
{\it i.e.} a positive output pulse if the input pulses are of opposite signs ($+ -$ or $- +$), 
and a negative output pulse if the input pulses are of equal signs ($+ +$ or $- -$).
Additionally, it should generate a null output if only one input pulse is received since the last go-cue signal.  
Variable time delays are introduced between the input pulses and the go-cues. 

A simpler version of the task was proposed 
in {\purple \cite{abbott2016building}}, whose {\green solution} involved 
first training an analog, rate-based neural network 
and replicating the learned network dynamics with a larger network of spiking neurons  ($\approx 3000$), using the method from predictive coding {\purple \cite{deneve2016efficient}}.
It also required a dendritic nonlinearity function 
to match the transfer function of rate neurons.

We trained a network of $80$ quadratic integrate and fire (QIF) neurons%
\footnote{NIF networks fail to learn the delayed-memory XOR task: the memory requirement for past input history 
drives the training toward strong recurrent connections and runaway excitation.}, 
whose dynamics is 
$$f(v,I) = (1 + \cos(2\pi v)) / \tau_v + (1 -  \cos(2\pi v)) I , $$ 
also known as Theta neuron model {\purple \cite{ermentrout2008ermentrout}}, with the threshold and the reset voltage at  $v_{\theta}=1$, $v_\text{reset}=0$. 
Time constants of $\tau_v=25$, $\tau_f=5$, and $\tau=20$ ms were used,
whereas the time-scale of the task was $\approx 500$ ms, much longer than the time constants. 
The intrinsic nonlinearity of the QIF spiking dynamics proves to be sufficient for solving this task without requiring extra dendritic nonlinearity. 
The trained network successfully solves  the delayed-memory XOR task (Figure 4): 
The spike patterns exhibit time-varying, but sustained activities that maintain the input history, 
generate the correct outputs when triggered by the go-cue signal,
and then return to the background activity. 
More analysis is needed to understand the exact underlying computational mechanism.

This result shows that out algorithm can indeed optimize spiking networks to perform nonlinear computations over extended time. 

\section{Discussion}

We have presented a novel, differentiable formulation of spiking neural networks and derived the gradient calculation for supervised learning. 
Unlike previous learning methods, our method 
optimizes the spiking network dynamics for general supervised tasks 
on the time scale of individual spikes as well as the behavioral time scales.

{\blue Exact gradient-based learning methods 
inevitably	involve discrepancies from biological learning processes.} 
Nonetheless, {\red such} methods provide solid theoretical ground for understanding the principles of biological learning rules.
For example, our result shows that the gradient update occurs in a sparsely compressed manner near spike times, bearing close resemblance to reward-modulated STDP.
{\blue Moreover, further analysis may reveal} 
that certain aspects of the gradient calculation can be approximated in a biologically plausible manner without significantly compromising the efficiency of optimization. 
For example, it was recently shown 
that the biologically implausible aspects of backpropagation method can be resolved through 
feedback alignment for rate-based multi-layer feedforward networks {\purple \cite{lillicrap2016random}}.
Such approximations could also apply to spiking neural networks. 


Here, we coupled the synaptic current model with 
differentiable single-state spiking neuron models.  
However, the synapse model 
can be coupled with any neuron models, 
including {\cyan realistic} multi-state neuron models with
detailed action potential dynamics%
\footnote{Simple modification of the gate function would be required to prevent activation during the falling phase of action potential.},
including the Hodgkin-Huxley model, the Morris-Lecar model, and the FitzHugh-Nagumo model;
and even models with internal adaptation variables.
It can also be coupled with neuron models having non-differentiable reset dynamics,
such as the leaky integrate and fire model, the exponential integrate and fire model,  
and the Izhikevich model,
although gradient calculation on these models would require additional procedures.
This will be examined in the future work.

\subsubsection*{Acknowledgments}
We thank Peter Dayan for helpful discussions.

\small

\bibliographystyle{unsrt}

\newpage 

\section*{Supplementary Materials: Gradient calculation for the spiking neural network}

\paragraph{Pontryagin's minimum principle}
According to {\purple \cite{pontryagin1962mathematical}},
the Hamiltonian for the spiking network dynamics eq~(\ref{eq:syn_dyn_new},\ref{eq:memV_dyn},\ref{eq:I}) is
\begin{align*}
	\mathcal{H} 	& =  \sum_i \bar{p}_{v_i}  \dot{v}_i  +  \bar{p}_{s_i}  \dot{s}_i + l(\vec{s}) \\
				& =  \sum_i (\bar{p}_{v_i}  +  g(v_i)  \bar{p}_{s_i}  /\tau ) f({v}_i,{I}_i)  -  (\bar{p}_{s_i}/\tau )  {s}_i + l(\vec{s}) ,
\end{align*}
where $\bar{p}_{v_i}$ and $ \bar{p}_{s_i}$ are the adjoint state variables for the membrane voltage $v_i$ and the synaptic current $s_i$ of neuron $i$, respectively, and $l(\vec{s})$ is the cost function. 
The back-propagating dynamics of the adjoint state variables are:
\begin{align*}
-\dot{\bar{p}}_{v_i} & = \frac{\partial \mathcal{H}}{\partial {v_i}} =  (\bar{p}_{v_i} +  g_i \bar{p}_{s_i}  /\tau ) f_{v_i}  +   f_{i} g_i' \bar{p}_{s_i}  /\tau \\
 -\dot{\bar{p}}_{s_i} & = \frac{\partial \mathcal{H}}{\partial {s_i}}
=    \sum_j  (\bar{p}_{v_j}  +  g_j \bar{p}_{s_j}  /\tau )\cdot f_{I_j}  W_{ji}    - \bar{p}_{s_i}/\tau + l_{s_i}  
\end{align*}
where
$f_v  \equiv \partial f/ \partial v$, $f_I  \equiv \partial f/ \partial I$, $g'  \equiv dg/dv$, and $ l_{s_i}  \equiv \partial l/ \partial s_i$.

This formulation can be simplified by  change of variables,
${p}_{v} \equiv \bar{p}_{v} +  g \bar{p}_{s} /\tau$,  $ {p}_{s}  \equiv \bar{p}_{s} /\tau$,
which yields
\begin{align*}
	\mathcal{H} 
				&  =  \vec{p}_{v} \cdot  \vec{f}  -  \vec{p}_{s}   \cdot\vec{s} + l  \\ 
	- \dot{p}_{v_i} 	& = f_{v_i}  {p}_{v_i}  - g_i   \dot{p}_{s_i} \\
        -  \tau   \dot{p}_{s_i} &  =  -  {p}_{s_i}  + l_{s_i} + \sum_j  W_{ji}  f_{I_j}   {p}_{v_j}   ,  
\end{align*}
where we used
$\dot{p}_{v_i} = \dot{\bar{p}}_{v_i} + f_i g_i' \bar{p}_{s_i} /\tau + g_i \dot{\bar{p}}_{s_i} /\tau$.

The gradient of the total cost can be obtained by integrating the partial derivative of the Hamiltonian with respect to the parameter
({\it e.g.}
$ \partial \mathcal{H}/\partial W_{ij}$,
$ \partial \mathcal{H}/\partial U_{ij}$,
$ \partial \mathcal{H}/\partial I_{o_i}$,
$ \partial \mathcal{H}/\partial O_{ij}$).

\end{document}